\documentclass[aps,prf,onecolumn,groupedaddress,superscriptaddress,notitlepage]{revtex4-1}\usepackage{natbib}
\usepackage{docs}
\usepackage{bm}
\usepackage[colorlinks=true,linkcolor=blue]{hyperref}
\usepackage{revsymb4-1}
\usepackage{graphicx}
\usepackage{natbib}
\usepackage{ulem}
\usepackage{amssymb}
\usepackage{amsmath,amsfonts}
\usepackage{ulem}
\usepackage{color}
\expandafter\ifx\csname package@font\endcsname\relax\else
\expandafter\expandafter
\expandafter\usepackage
\expandafter\expandafter
\expandafter{\csname package@font\endcsname}
\fi
\begin{document}
\title{Cooperative strings and glassy dynamics in various confined geometries}
\author{Maxence Arutkin}
\affiliation{Perimeter Institute for Theoretical Physics, 31 Caroline St N, Waterloo, Ontario, Canada.}
\affiliation{UMR CNRS Gulliver 7083, ESPCI Paris, PSL Research University, 75005 Paris, France.}
\author{Elie Rapha\"{e}l}
\affiliation{UMR CNRS Gulliver 7083, ESPCI Paris, PSL Research University, 75005 Paris, France.}
\author{James A. Forrest}
\affiliation{Perimeter Institute for Theoretical Physics, 31 Caroline St N, Waterloo, Ontario, Canada.}
\affiliation{UMR CNRS Gulliver 7083, ESPCI Paris, PSL Research University, 75005 Paris, France.}
\affiliation{Department of Physics \& Astronomy and Guelph-Waterloo Physics Institute, University of Waterloo, Waterloo, Ontario N2L 3G1, Canada.}
\author{Thomas Salez}
\email{thomas.salez@u-bordeaux.fr}
\affiliation{Perimeter Institute for Theoretical Physics, 31 Caroline St N, Waterloo, Ontario, Canada.}
\affiliation{Univ. Bordeaux, CNRS, LOMA, UMR 5798, F-33405 Talence, France.}
\affiliation{Global Station for Soft Matter, Global Institution for Collaborative Research and Education, Hokkaido University, Sapporo, Japan.}
\begin{abstract}
Previously, we developed a minimal model based on random cooperative strings for the relaxation of supercooled liquids in the bulk and near free interfaces, and we recovered some key experimental observations. In this article, after recalling the main ingredients of the cooperative string model, we study the effective glass transition and surface mobility of various experimentally-relevant confined geometries: freestanding films, supported films, spherical particles, and cylindrical particles, with free interfaces and/or passive substrates. Finally, we introduce a novel way to account for a purely attractive substrate, and explore the impact of the latter in the previous geometries.
\end{abstract}
\maketitle

\section{Introduction}
Understanding the nature of the glass transition and the physical properties of glasses remains a key open problem in condensed matter physics. The observed dynamics of a glass-forming liquid close to the glass transition temperature follows the Vogel-Fulcher-Tammann phenomenology, for which a low-temperature extrapolation of the viscosity \textit{vs} temperature data leads to an apparent divergence of the viscosity at finite temperature~\cite{vogel1921law,fulcher1925analysis,tammann1926abhangigkeit}. A significant effort has been put forward into developing and improving theoretical descriptions of the fascinating properties of glass-forming materials, and understanding glass formation. The fact that some of these theories are thermodynamic in nature - predicting \textit{e.g.} an underlying real phase transition - and others are purely kinetic provides an indication on just how rich the phenomenology is. There is a number of things that are known to be important physical ingredients of glass formation. One idea, for instance, is that the energy landscape of glassy systems contains many local minima, and transitions between these minima controls the dynamics. Recently, it has been shown that an energy landscape approach can provide a solvable model of glasses~\cite{Charbonneau2017}.

Another feature of glassy dynamics which was proven to be robust is the idea of cooperative motion. The original scenario, proposed by Adam and Gibbs~\cite{adam1965temperature}, suggests that relaxation in glassy systems involves collective molecular rearrangements that occur within independent cooperative clusters. The length scale $\xi$ of these cooperative regions has raised an important interest~\cite{donth1996characteristic}, and has triggered the investigation of glassy dynamics in finite-sized and confined systems~\cite{jackson1990melting}. Glassy polymer nanofilms~\cite{forrest2001glass} are probably the most prominent examples of the latter, and for the emblematic case of polystyrene there is a considerable body of experimental evidence supporting the fact that the glass transition temperature decreases as the film thickness is reduced below several tens of nanometers, when there is at least one free interface. Interestingly, the same behavior has been observed for nanopores~\cite{jackson1990melting}, and spherical nanoparticles~\cite{feng2013glass,rharbi2008reduction,zhang2011glass}, among other geometries. Besides, experiments and numerical studies have shown that the mobility of glassy materials is higher near a free interface compared to the bulk~\cite{fakhraai2008measuring,Malshe2011,Hoang2011,ediger2013dynamics,yang2010glass,chai2014direct,yoon2014substrate,Kuon2018,Tanis2019,Chai2019}.

As a consequence, it appears that there would be a clear benefit to having a predictive model of glassy dynamics, which could be implemented in a straightforward manner for finite-sized and confined systems, in any of the specific experimental geometries employed or envisioned. Such a description should also be able to characterize the effects of interfacial interactions on the predicted behaviors. In previous studies, we developed a minimal model based on cooperative strings for the relaxation of supercooled liquids near free interfaces~\cite{salez2015cooperative}, and by applying it to thin films~\cite{salez2015cooperative}, and spherical nanoparticles~\cite{arutkin2016cooperative}, we recovered the key experimental observations. In this article, after recalling the main ingredients of the cooperative string model in the bulk and near free interfaces, we study the glass transition and surface mobility of various experimentally-relevant confined geometries involving free interfaces. Finally, we introduce a novel way to account for a purely attractive substrate in the model, and explore the impact of the latter in the previous geometries.

\section{Cooperative string model}
\subsection{Bulk description}
Adam and Gibbs introduced the key idea that, in a supercooled liquid, a local relaxation event results from a cooperative sequence of individual liquid-like motions~\cite{adam1965temperature}. The number $N^*$ of particles typically involved in such a local relaxation is called the cooperativity, and it depends on the average density $\phi$ of the supercooled liquid. Essentially, the denser the system is, the larger the cooperativity has to be due to increased free-volume constraints. Based on numerical~\cite{Donati1998,Stevenson2006,Betancourt2014} and experimental~\cite{Zhang2011} observations reported in the literature, and an argument of analytical simplicity, the cooperative string model~\cite{salez2015cooperative} further idealizes a cooperative region into a random train of molecules temporarily squeezing against each other in order to free enough space for a rearrangement to occur. From that picture, and invoking thermal expansion, the rate law for the relaxation process self-consistently results in the Vogel-Fulcher-Tammann time-temperature superposition. Below, we briefly recall the main ingredients of that description.

We introduce a coherence probability $\epsilon$, \textit{i.e.} a typical liquid-like Boltzmann probability for a given molecule to reach a certain energy barrier resulting from the interaction with a nearest neighbour. The probability $\mathcal{P}$ of a cooperative relaxation involving $N^*$ independent but coherent molecular motions then reads:
\begin{equation}
\mathcal{P}\sim \epsilon^{N^*}\ .
\label{pn}
\end{equation}
Introducing a molecular time scale $\tau_0$, the liquid-like time scale $\tau_{\textrm{c}}$, and invoking ergodicity through $\mathcal{P}\sim\tau_0/\tau$ and $\epsilon\sim\tau_0/\tau_{\textrm{c}}$, the relaxation time $\tau$ in the supercooled liquid follows the Adam-Gibbs phenomenology:
\begin{equation}
\frac{\tau}{\tau_0}\sim \left(\frac{\tau_{\textrm{c}}}{\tau_0}\right)^{N^*}\ .
\label{AG}
\end{equation}

Using mean-field arguments, the cooperativity can further be linked to the density through:
\begin{equation}
N^{*}(\phi)\sim \frac{\phi_{\textrm{V}}^{\,1/3}-\phi_{\textrm{c}}^{\,1/3}}{\phi_{\textrm{V}}^{\,1/3}-\phi^{1/3}}\ ,
\label{cooperativity}
\end{equation}
where $\phi_{\textrm{c}}$ and $\phi_{\textrm{V}}$ are respectively the densities at the onset of cooperativity ($N^*\sim1$) and at the kinetic arrest point ($N^*\rightarrow+\infty$). Within this simplified picture, the temperature $T$ essentially arises in the effect it has on the material density. In the relevant density range where $\phi\in[\phi_{\textrm{c}};\phi_{\textrm{V}}]$, the material dilatation is small in practice, so that $\phi(T)\simeq\phi_{\textrm{V}}[1+\alpha(T_{\textrm{V}}-T)]$, where $\alpha$ is the thermal expansion coefficient, and $T_{\textrm{V}}$ is the Vogel temperature identified as the temperature at the kinetic arrest point. 

Combining that linear thermal expansion with Eqs.~(\ref{AG}) and~(\ref{cooperativity}), one obtains the Vogel-Fulcher-Tamman law at leading order near the kinetic arrest point:
\begin{equation}
\tau(T)\sim \tau_0\,\exp\left(\frac{A}{T-T_{\textrm{V}}}\right) \ ,
\end{equation}
where $A\sim(T_{\textrm{V}}-T_{\textrm{c}})\ln(\epsilon)$ is a reference temperature, and $T_{\textrm{c}}$ is the cooperative onset temperature corresponding to the density $\phi_{\textrm{c}}$. With this time-temperature superposition, the so-called bulk glass transition temperature $T_{\textrm{g}}^{\text{bulk}}$ simply corresponds to a specific, long (say hundreds of seconds), experimental time scale $\tau_{\textrm{g}}\sim\tau(T_{\textrm{g}}^{\text{bulk}})$. 

Finally, the cooperative strings are essentially random walks. Consequently, the length scale $\xi$ of a cooperative region typically reads $\xi\sim\lambda_{\textrm{V}} \sqrt{N^*}$, and thus:
\begin{equation}
\xi(T)\sim\lambda_{\textrm{V}}\sqrt{\frac{T_{\textrm{c}}-T_{\textrm{V}}}{T-T_{\textrm{V}}}} \ , 
\label{xi_formula}
\end{equation}
where $\lambda_{\textrm{V}}$ is the typical interparticle distance at kinetic arrest (\textit{i.e} a molecular size), and where the exponent 1/2 reflects the ideal random walk assumption, which should be modified for more realistic walks. 

\subsection{Free interface}
\begin{figure}[t!]
\centering
\includegraphics[scale=0.3]{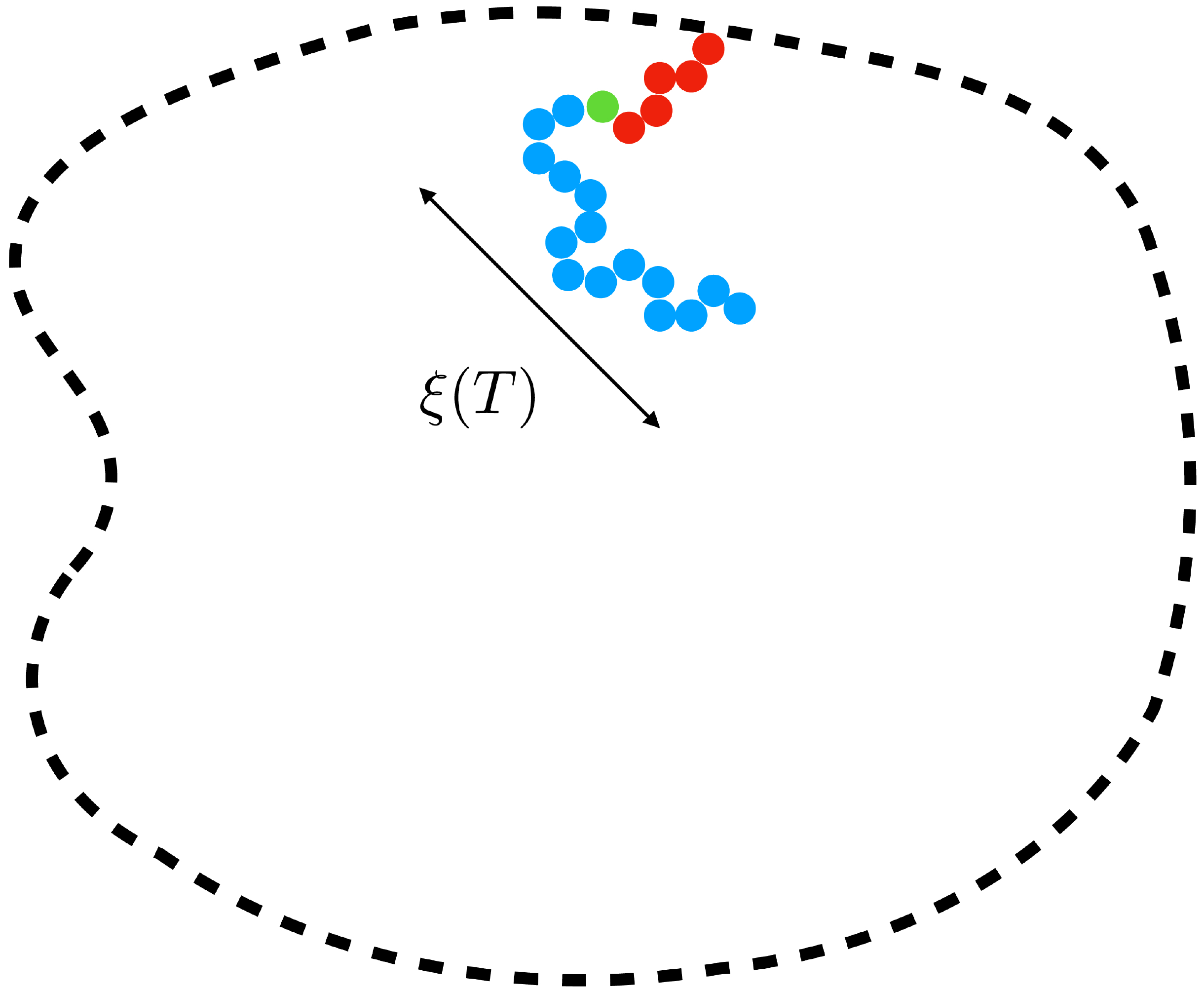}
\caption{Random cooperative strings inside a supercooled liquid whose volume is delimited by an interface (dashed line). The relaxation of a test particle (green) can occur through a bulk cooperative string (blue) of size $\xi$. In the case of a free interface, it can also occur through a shorter string (red) due to truncation by the interface (\textit{i.e.} absorbing boundary condition). If the interface is a passive substrate, it simply reflects the string (\textit{i.e.} reflecting boundary condition). Finally, if the interface is capped by a purely attractive substrate (\textit{i.e.} purely attractive boundary condition), the shorter string (red) does not contribute to relaxation, and another random attempt must be made from the origin (green).}
\label{Fig1}
\end{figure}
The effect of a free interface is included in our model through the truncation of any cooperative string reaching it (see Fig.~\ref{Fig1})~\cite{salez2015cooperative}. Indeed, such an interface represents an infinite reservoir of free volume, and thus, as soon as a string touches it, no additional cooperative neighbour is needed in order to free the missing space for relaxation. This truncation mechanism then naturally leads to an average local cooperativity, defined at position $\bm{r}$ as:
\begin{equation}
N_{\textrm{s}}^*(\bm{r})=\langle\min\left( N^*,n_0\right)\rangle =N^*f\left(\frac{\bm{r}}{\xi}\right)\ ,
\label{defmin}
\end{equation}
where the random variable $n_0$ is the number of molecules in a cooperative string between its starting point $\bm{r}$ and its first passage at the interface, $\langle\rangle$ is the average over $n_0$, and where we introduced the cooperative reduction factor $f=N_{\textrm{s}}^*/N^*$ and its natural dimensionless variable $\bm{r}/\xi$. We further stress that the probability distribution of $n_0$, and thus $f$, depend on the location of the boundary points with respect to the starting point $\bm{r}$, which will be crucial later when addressing various specific geometries. By construction, at a given temperature $T$, the local cooperativity $N_{\textrm{s}}^*$ is always smaller than the bulk one $N^*$. When the distance to the interface is large compared with the bulk cooperative length scale $\xi$, the free interface is typically not reached with less than $N^*$ cooperative neighbours, and thus the behaviour is bulk-like with $N_{\textrm{s}}^*\sim N^*$ or $f\sim1$. In contrast, when the distance to the interface vanishes, the free-volume constraints disappear, and thus the behaviour is molecular-like (or gaseous-like) with $N_{\textrm{s}}^*\rightarrow0$ or $f\rightarrow0$. 

Within this framework, the local relaxation time $\tau_{\textrm{s}}$ in a supercooled liquid with a free interface is essentially obtained by replacing $N^*$ by $N_{\textrm{s}}^*$ in Eq.~(\ref{AG}), which leads to:
\begin{equation}
\frac{\tau_{\textrm{s}}(\bm{r})}{\tau_0}\sim \left(\frac{\tau}{\tau_0}\right)^{f\left(\frac{\bm{r}}{\xi}\right)} \ .
\label{exAG}
\end{equation}
This formula extends the bulk Adam-Gibbs phenomenology to cases in which a free interface is present. Interestingly, the cooperative reduction factor $f$ acts as a novel exponent on the reference bulk law, with values ranging between 0 and 1 depending on the distance to the interface. It fully determines how the free interface enhances the dynamics. 

\subsection{Diffusion framework}
In the limit of vanishing spatial step size $\lambda_{\textrm{V}}$ and infinite number of steps, we assimilate the ideal random walks modelling the cooperative strings to Brownian trajectories, and we invoke a diffusion framework in order to describe their statistics. This description is asymptotically valid near the kinetic arrest point, where the bulk cooperativity is large. While the approximation is certainly less valid for smaller strings, it is hoped as in critical phenomena that one may still extract some important characteristic features further away from the divergence point. 

We thus introduce the two relevant rescaled variables: the starting position $\bm{R}=\bm{r}/\xi$, and the curvilinear abscissa $l=n/N^*$ along the string, where $n$ is the $n$-th molecule in the string. In the presence of a free interface located at the boundary $\partial D$ of the domain $D$ (see Fig.~\ref{Fig1}), the probability density $P(\bm{R'},l\vert\bm{R})$ for a string starting at position $\bm{R}$ in the domain $D$ to reach the position $\bm{R'}$ in the domain $D$, at a curvilinear abscissa $l$, satisfies:
\begin{align}
\label{EqDiff0}
&\partial_l P=\frac{1}{2}\Delta P\ ,\\
&P(\bm{R'},0\vert\bm{R})=\delta(\bm{R}-\bm{R'})\ ,\\
&P(\bm{R'},l\vert \bm{R})=0 \ , \ \forall \bm{R'}\in \partial D \ ,
\label{EqDiff}
\end{align}
where the Laplace operator $\Delta$ acts on the variable $\bm{R'}$. This set of equations corresponds to a diffusion equation in a bounded Euclidean domain, with a Dirac initial condition and an absorbing (\textit{i.e.} truncating) Dirichlet boundary condition. Note that, to describe a passive solid wall rather than a free interface, one would introduce instead a reflecting Neumann boundary condition: $\partial_n P(\bm{R'},l\vert \bm{R})=0 \ , \ \forall \bm{R'}\in \partial D$, where $\partial_n$ is the derivative normal to the boundary and pointing outside the domain. 

No matter the chosen boundary condition (including a mixture of the two above), the spectrum of the Laplace operator in such a bounded domain is discrete with the following general properties~\cite{courant1966methods,kac1966can,grebenkov2013geometrical}. First, the eigenvalues $-2\lambda_n$ are real, negative, with $\lambda_n$ increasing with the integer $n$. Secondly, the corresponding eigenfunctions $u_n(\bm{R})$ form a basis on $\textrm{L}^2(D)$, and satisfy the boundary condition. Thirdly, the set of eigenfunctions is complete:
\begin{equation}
\sum_{n=1}^{+\infty}u_n(\bm{R})u_n^*(\bm{R'})=\delta(\bm{R}-\bm{R'}) \ .
\end{equation}
Finally, the eigenfunctions are orthonormal:
\begin{equation}
\int_D\text{d}\bm{R}\,u_n(\bm{R})u_m^*(\bm{R})=\delta_{nm} \ .
\end{equation} 
By unicity, $P(\bm{R'},l\vert \bm{R})$ is the heat kernel of Eqs.~(\ref{EqDiff0}-\ref{EqDiff}), which reads:
\begin{equation}
P(\bm{R'},l\vert \bm{R})=\sum_{n}u_n(\bm{R})u_n^*(\bm{R'})e^{-\lambda_n l} \ .
\end{equation}

For the case of an absorbing condition at the boundary (or part of it), by integrating $P(\bm{R'},l\vert \bm{R})$ over all possible arrival points $\bm{R'}$ within the domain $D$, one obtains the survival probability $S({\bm{R}},l)$, \textit{i.e.} the probability for a string not to encounter the absorbing boundary before the curvilinear abscissa $l$. It reads:
\begin{equation}
\label{survival}
S({\bm{R}},l)=V\sum_{n}a_n u_n(\bm{R})e^{-\lambda_n l} \ ,
\end{equation}
with $V=\int_D\text{d}\bm{R'}$, and $a_n=[\int_D\text{d}\bm{R'}\,u_n^*(\bm{R'})]/V$. Since, in turn, $1-S({\bm{R}},l)$ represents the probability for a string to encounter the absorbing boundary for the first time before the curvilinear abscissa $l$, the probability density for a string to encounter the absorbing boundary for the first time at the curvilinear abscissa $l_0=n_0/N^*$ is $g(\bm{R},l_0)=-\partial_lS({\bm{R}},l)\vert_{l=l_0}$. The latter quantity is the so-called first-passage density and it thus reads~\cite{redner2001guide}:
\begin{equation}
\label{fpd}
g(\bm{R},l_0)=V\sum_{n}\lambda_n a_n u_n(\bm{R}) e^{-\lambda_n l_0}\ .
\end{equation}
Finally, invoking Eq.~(\ref{defmin}), the cooperative reduction factor in presence of a free interface is defined as $f(\bm{R})=\int_0^{+\infty}\textrm{d}l_0\,g(\bm{R},l_0)\min(1,l_0)$, and it thus has the general form:
\begin{equation}
f(\bm{R})=V\sum_{n}a_n u_n(\bm{R})\,\frac{1- e^{-\lambda_n}}{\lambda_n} \ .
\label{coopred}
\end{equation}
As shown below, for simple enough geometries in which symmetries allow for separation of variables, and thus explicit representations of the spectrum $\{\lambda_n\}$ and eigenfunctions $\{u_n(\bm{R})\}$, this central quantity of the cooperative string model near a free interface can be calculated. 

\subsection{Physical observables}
\label{princ}
An important feature of the cooperative string model, is the practical ability it provides to calculate physical observables for the glass transition in confinement and at interfaces. Among those, two central quantities have caught an intense scientific attention: the mobile layer thickness $h_{\textrm{m}}$ and the effective glass transition temperature $T_{\textrm{g}}$. For a sufficiently simple geometry involving absorbing (free interface) and/or reflecting (passive substrate) boundaries, one can in principle calculate (or estimate numerically through a truncation of the infinite sum involved) the cooperative reduction factor from Eq.~(\ref{coopred}). Inserting it in Eq.~(\ref{exAG}) and recalling that $\bm{R}=\bm{r}/\xi$, one obtains the local relaxation time $\tau_{\textrm{s}}$ for any position $\bm{r}$ in the system, at the considered temperature $T$. Invoking the above typical time scale $\tau_{\textrm{g}}$ defining the glass transition, one can separate at any given temperature the liquid regions of the system for which $\tau_{\textrm{s}}<\tau_{\textrm{g}}$, from the glassy regions for which $\tau_{\textrm{s}}>\tau_{\textrm{g}}$. Note that each of these two phases is expected to have its own physical (\textit{i.e.} calorimetric, dilatometric, optical, etc.) properties.

From that phase separation, one can then do two things. First, as the temperature is decreased below $T_{\textrm{g}}^{\text{bulk}}$, the fraction of glassy regions grows. As a consequence, there might typically exist a temperature at which the system contains equal volumes of glassy and liquid regions, and thus exhibits average physical properties between the two phases. This temperature is identified to the effective glass transition temperature $T_{\textrm{g}}$ probed in global measurements. For simple enough geometries, $T_{\textrm{g}}$ can be computed as a function of the main size (\textit{e.g.} thickness, radius) of the system, or the fraction of free interfaces (\textit{e.g.} density of inclusions in a composite material). Secondly, at any temperature $T$ below $T_{\textrm{g}}^{\text{bulk}}$, the liquid regions typically localize near the free interfaces and form the so-called mobile layer. For simple enough geometries exhibiting certain symmetries (\textit{e.g.} translational or rotational invariance), this mobile layer may be fully characterized by a single length: its thickness $h_{\textrm{m}}$, that can thus be computed and plotted as a function of temperature. 

As a final remark, since the three (note that $A$ self-cancels with the $\tau_{\textrm{s}}\sim\tau_{\textrm{g}}$ criterion above) parameters $T_{\textrm{g}}^{\text{bulk}}$, $T_{\textrm{V}}$, and $T_{\textrm{c}}$, can be \textit{a priori} determined from bulk experiments on a given material, the only unknown quantity in the cooperative string model for confinement studies with the same material is the (molecular) size $\lambda_{\textrm{V}}$. The latter can be \textit{e.g.} obtained for a given geometry though a fit of the effective glass transition temperature data \textit{vs} system size~\cite{salez2015cooperative}, and then further used to calculate the mobile layer thickness $h_{\textrm{m}}$ as a function of temperature $T$ for various system sizes, with no free parameter. For the purpose of the present study, we invoke the values for (low molecular weight) polystyrene reported previously: $T^{\textrm{bulk}}_{\textrm{g}} = 371\ \textrm{K}$~\cite{Rubinstein2003}, $T_{\textrm{V}}=322\ \textrm{K}$~\cite{salez2015cooperative} (close to the established value $T_{\textrm{V}}=327\ \textrm{K}$~\cite{Sahnoune1996}), $T_{\textrm{c}}=463\ \textrm{K}$~\cite{Donth1996,Kahle1997}, and $\lambda_{\textrm{V}}=3.7$~nm~\cite{salez2015cooperative}.

\section{Application to various geometries}
\subsection{Freestanding films}
\begin{figure}[!t]
\centering
\includegraphics[scale=0.4]{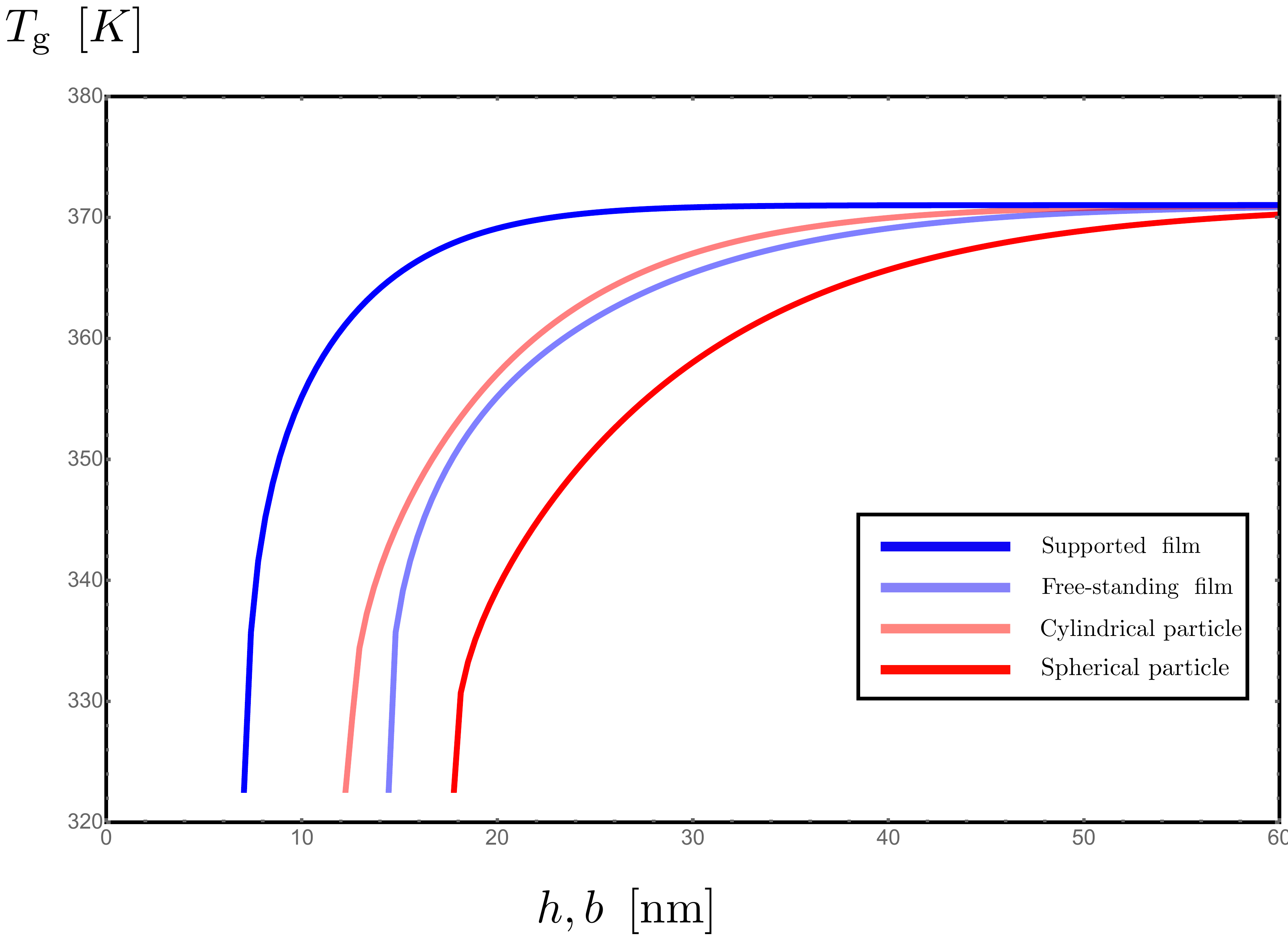}
\caption{Effective glass transition temperature of (low molecular weight) polystyrene as a function of system size (\textit{i.e.} film thickness $h$ or particle radius $b$), for various confined geometries involving free interfaces, and in one case (supported film) also a passive substrate, as computed from Eqs.~(\ref{solfree}),~(\ref{solsup}),~(\ref{solsphere}), and~(\ref{solcyl}), as well as the procedure detailed in Sec.~\ref{princ}.}
\label{Fig2}
\end{figure}
We describe here a freestanding film of thickness $h$ as an infinite volume of material delimited by two parallel and flat free interfaces (\textit{i.e.} absorbing boundary conditions), located at $z=0$ and $z=h$ respectively, where $z$ is the coordinate of $\bm{r}$ along the axis perpendicular to the interfaces, the origin of $\bm{r}$ being chosen on one of the free interfaces. We introduce the rescaled quantities $Z=z/\xi$ and $H=h/\xi$, and the dimensionless free surface to volume ratio is $2/H$. For such a geometry, the separation of variables in the diffusion problem yields a unidimensional description along $Z$, with:
\begin{equation}
u_n(Z)=\sqrt{\frac{2}{H}}\sin\left(\omega_n Z\right) \  ,
\end{equation}
and:
\begin{equation}
\lambda_n=\frac{\omega_n^{\,2}}{2}=\frac{\pi^2n^2}{2H^2}\ , \ \forall n\geq1\ .
\end{equation}
Note that the integrals $\int_D\text{d}\bm{R}$ are replaced by $\int_0^H\text{d}Z$. With those specific spectral quantities, the cooperative reduction factor of Eq.~(\ref{coopred}) becomes:
\begin{equation}
\label{solfree}
f(Z,H)=\frac{8}{H}\sum_{k=0}^{+\infty}\sin\left(\omega_{2k+1} Z\right)\,\frac{1- e^{-\frac{\omega_{2k+1}^{\,2}}{2}}}{\omega_{2k+1}^{\,3}}\ ,
\end{equation}
where we have now indicated explicitly the previously hidden dependency on the geometry, through the unique confinement parameter $H$. 

\begin{figure}[!t]
\centering
\includegraphics[scale=0.4]{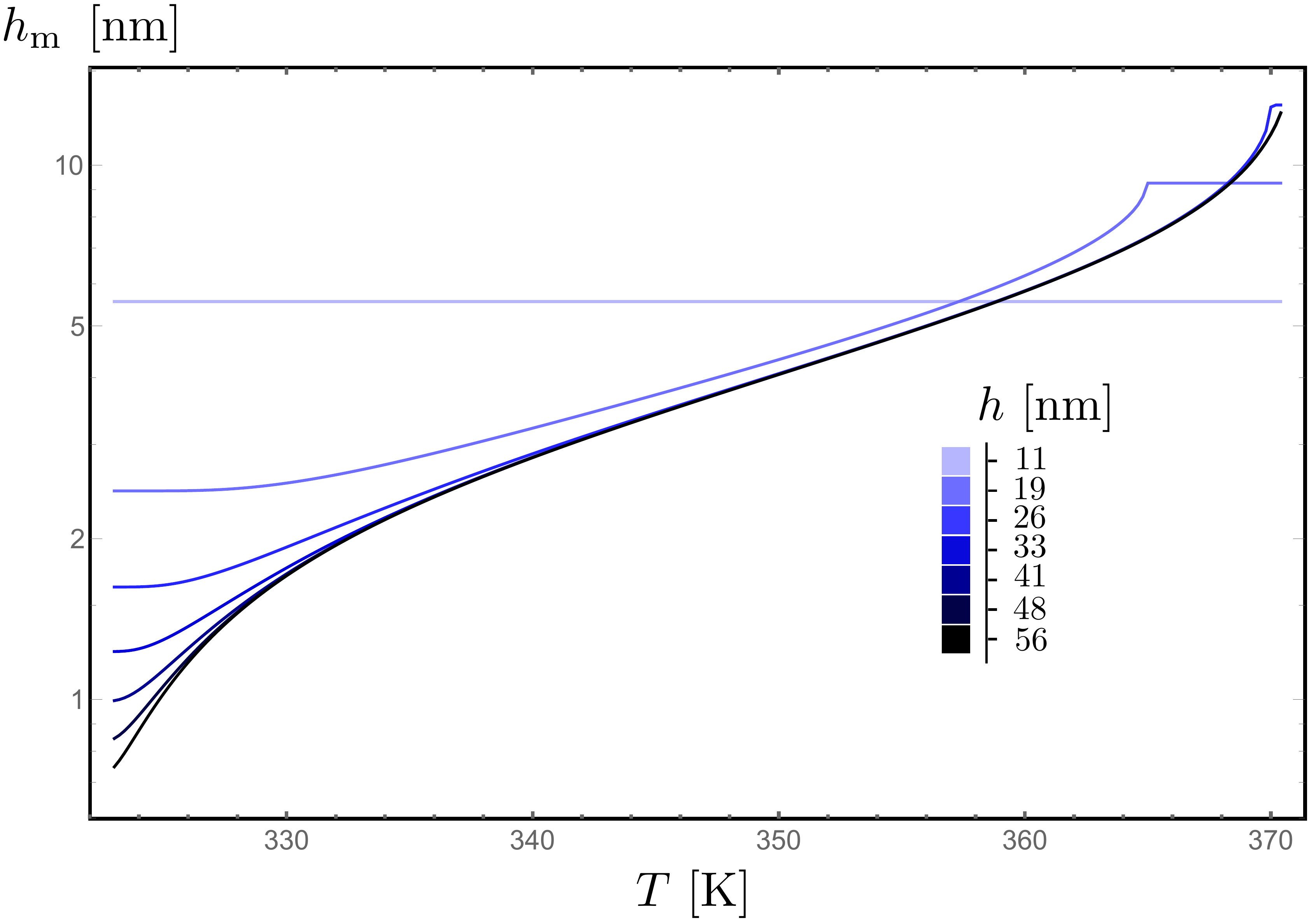}
\caption{Mobile layer thickness of (low molecular weight) polystyrene freestanding films as a function of temperature for several film thicknesses, as computed from Eq.~(\ref{solfree}) and the procedure detailed in Sec.~\ref{princ}. The corresponding plot for (low molecular weight) polystyrene supported films is simply obtained by replacing the label ``$h$" in the legend by ``$2h$", according to Eq.~(\ref{solsup}).}
\label{Fig3}
\end{figure}

Using Eq.~(\ref{solfree}) and the procedure detailed in Sec.~\ref{princ}, we can now compute $T_{\textrm{g}}(h)$ and $h_{\textrm{m}}(T,h)$ for (low molecular weight) polystyrene, which are plotted on Figs.~\ref{Fig2} and~\ref{Fig3}, respectively.
As predicted from the model, the effective glass transition temperature is reduced with respect to the bulk value for thicknesses below $\sim50$~nm, and this reduction reaches several tens of degrees in the considered thickness range. Furthermore, the predicted mobile layer thickness is nanometric, larger for thinner films, and it increases with temperature until it reaches a plateau at $h/2$ -- a situation corresponding to a pure liquid freestanding film since there is a mobile layer on each of the two sides of the film. Interestingly, from the model, below a critical thickness on the order of $\sim10$~nm, and in the considered temperature range between $T_{\textrm{V}}=322\ \textrm{K}$ and $T^{\textrm{bulk}}_{\textrm{g}} = 371\ \textrm{K}$, a (low molecular weight) polystyrene freestanding film should remain purely liquid and should not experience an effective glass transition. 

\subsection{Supported films}
In a previous work, we partly described a film supported on a passive substrate as a semi-infinite medium with one free interface~\cite{salez2015cooperative}. Here, we refine this description by considering instead an infinite volume of material delimited by two parallel and flat interfaces: one free interface (\textit{i.e.} absorbing boundary condition) located at $z=0$, and one passive wall (\textit{i.e.} reflecting boundary condition) located at $z=h$, where $h$ is the film thickness and $z$ is the coordinate of $\bm{r}$ along the axis perpendicular to the interfaces, the origin of $\bm{r}$ being chosen on the free interface. We introduce the rescaled quantities $Z=z/\xi$ and $H=h/\xi$, and the dimensionless free surface to volume ratio is $1/H$. For such a geometry, the separation of variables in the diffusion problem yields a unidimensional description along $Z$, with:
\begin{equation}
u_n(Z)=\sqrt{\frac{2}{H}}\sin\left(\omega_n Z\right) \  ,
\end{equation}
and:
\begin{equation}
\lambda_n=\frac{\omega_n^{\,2}}{2}=\frac{\pi^2(2n+1)^2}{8H^2}\ , \ \forall n\geq 0\ .
\end{equation}
Note that the integrals $\int_D\text{d}\bm{R}$ are replaced by $\int_0^H\text{d}Z$. With those specific spectral quantities, the cooperative reduction factor of Eq.~(\ref{coopred}) becomes:
\begin{equation}
\label{solsup}
f(Z,H)=\frac{4}{H}\sum_{n=0}^{+\infty}\sin(\omega_n Z)\,\frac{1- e^{-\frac{\omega_n^{\,2}}{2}}}{\omega_n^{\,3}}\ , 
\end{equation}
where we have now indicated explicitly the previously hidden dependency on the geometry, through the unique confinement parameter $H$. 

Using Eq.~(\ref{solsup}) and the procedure detailed in Sec.~\ref{princ}, we can now compute $T_{\textrm{g}}(h)$ for (low molecular weight) polystyrene. As seen in Fig.~\ref{Fig2}, there seems to be a simple homothetic relation between the curves for freestanding and supported films. Indeed, when performing the transformation $H\rightarrow H/2$ in Eq.~(\ref{solsup}), one obtains exactly Eq.~(\ref{solfree}). The plot of $h_{\textrm{m}}(T,h)$ for supported films is thus also simply obtained from Fig.~\ref{Fig3} by replacing the label ``$h$" in the legend by ``$2h$". This general mapping is naturally expected due to the reflection properties of Brownian motion, but it has a profound physical implication within the framework of the cooperative string model: a film supported on a passive substrate is indistinguishable from a freestanding film with double thickness, as far as the effective glass transition and glassy dynamics are concerned. Therefore, the model allows to recover the central experimental observation of Ref.~\cite{Mattsson2000}. 
\begin{figure}[!t]
\centering
\includegraphics[scale=0.4]{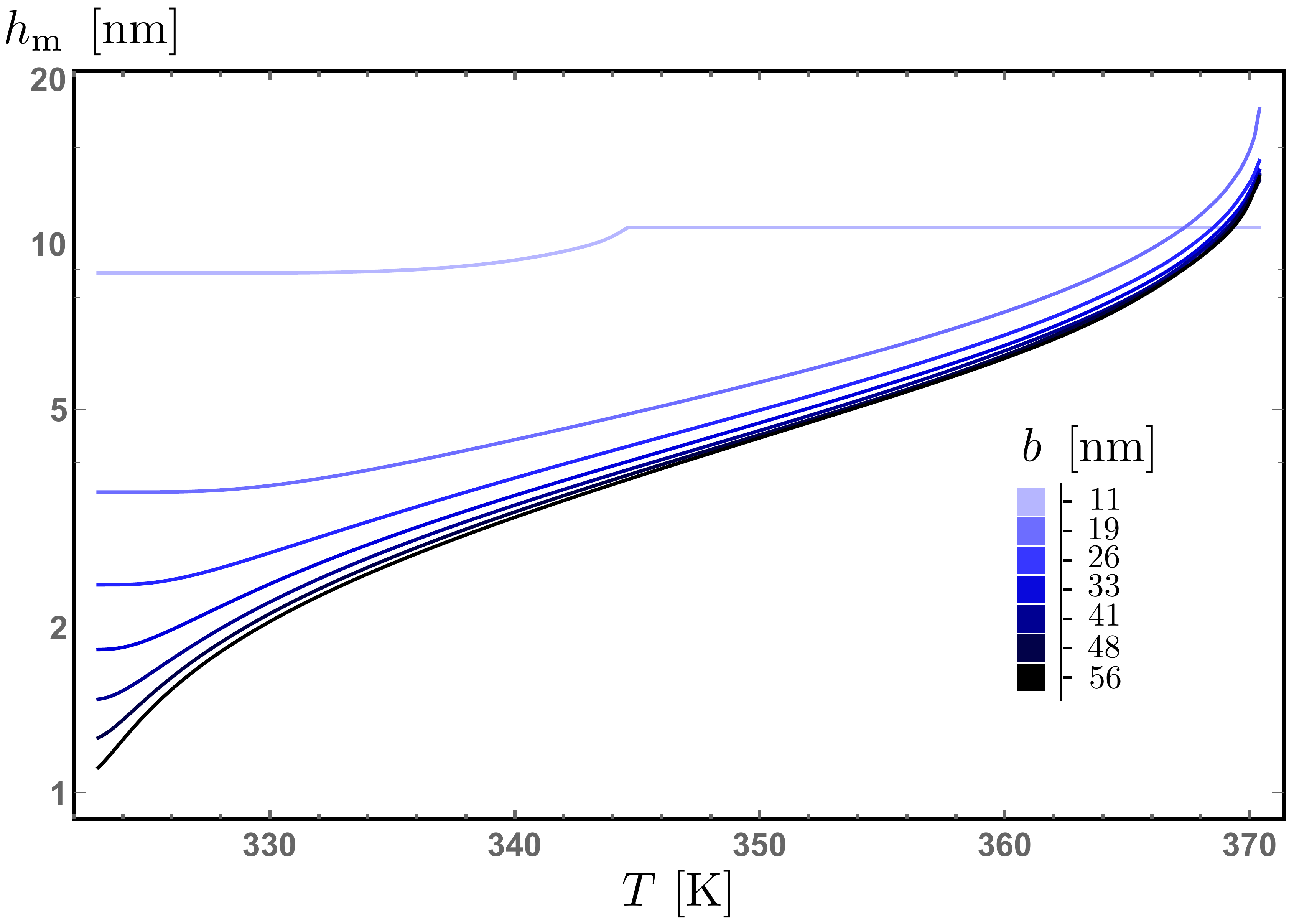}
\caption{Mobile layer thickness of (low molecular weight) polystyrene spherical particles as a function of temperature for several particle radii, as computed from Eq.~(\ref{solsphere}) and the procedure detailed in Sec.~\ref{princ}.}
\label{Fig4}
\end{figure}

\subsection{Spherical particles}
We consider here a sphere with radius $b$, delimited by a free interface (\textit{i.e.} absorbing boundary condition) located at $r=b$, where $r$ is the norm of $\bm{r}$, the origin of $\bm{r}$ being chosen at the center of the sphere. We introduce the rescaled quantities $R=r/\xi$ and $B=b/\xi$, and the dimensionless free surface to volume ratio is $3/B$. For such a geometry, the separation of variables in the diffusion problem yields a unidimensional description along $R$, with:
\begin{equation}
u_n(R)=\sqrt{\frac{2}{B}} \frac{\sin\left(\omega_n R\right)}{R} \  ,
\end{equation}
and:
\begin{equation}
\lambda_n=\frac{\omega_n^{\,2}}{2}=\frac{\pi^2n^2}{2B^2}\ , \ \forall n\geq1\ .
\end{equation}
Note that the integrals $\int_D\text{d}\bm{R}$ are replaced by $\int_0^B\text{d}R\,R^2$. With those specific spectral quantities, the cooperative reduction factor of Eq.~(\ref{coopred}) becomes:
\begin{equation}
\label{solsphere}
f(R,B)=\frac{4}{R}\sum_{n=1}^{+\infty}(-1)^{n-1}\sin(\omega_n R)\,\frac{1-e^{-\frac{\omega_n^{\,2}}{2}}}{\omega_n^{\,3}}\ ,
\end{equation}
where we have now indicated explicitly the previously hidden dependency on the geometry, through the unique confinement parameter $B$. Note that we recover here the result obtained in a previous work with a different method~\cite{arutkin2016cooperative}.
\begin{figure}[!t]
\centering
\includegraphics[scale=0.4]{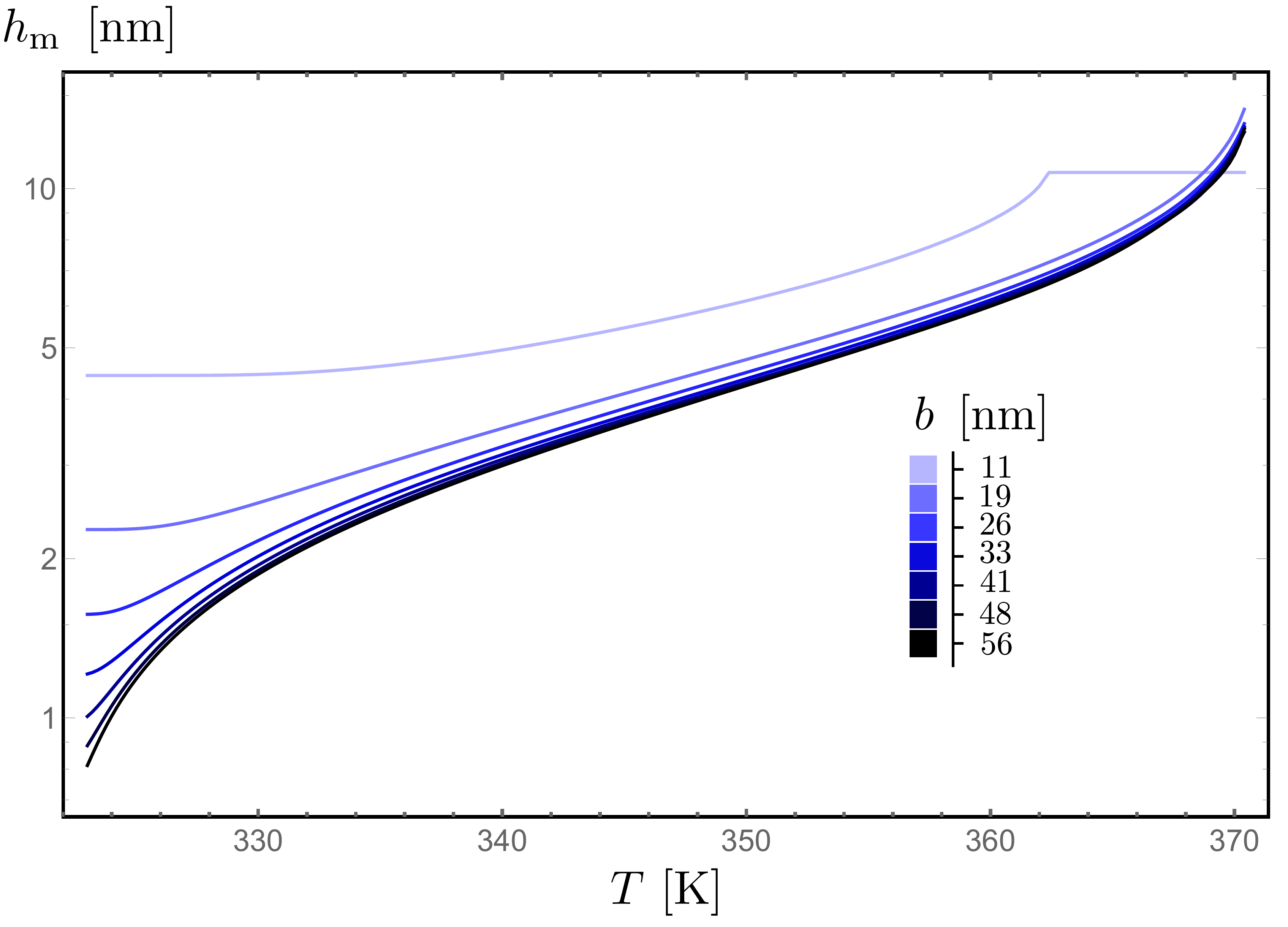}
\caption{Mobile layer thickness of (low molecular weight) polystyrene cylindrical particles as a function of temperature for several particle radii, as computed from Eq.~(\ref{solcyl}) and the procedure detailed in Sec.~\ref{princ}.}
\label{Fig5}
\end{figure}

Using Eq.~(\ref{solsphere}) and the procedure detailed in Sec.~\ref{princ}, we can now compute $T_{\textrm{g}}(b)$ and $h_{\textrm{m}}(T,b)$ for (low molecular weight) polystyrene, which are plotted on Figs.~\ref{Fig2} and~\ref{Fig4}, respectively. As for the other geometries, the predicted mobile layer thickness is nanometric, larger for smaller spherical particles, and it increases with temperature until it reaches a plateau at $b$ -- a situation corresponding to a pure liquid spherical particle. Besides, among the four presented geometries, the sphere is the one that maximizes the free surface to volume ratio, for identical $b$ and $h$ values. Therefore, as expected, while the $T_{\textrm{g}}(b)$ curve for spherical particles is qualitatively similar to the corresponding curves for other geometries, the effect of the free interface is more pronounced for spherical particles. 

\subsection{Cylindrical particles}
We describe here a cylindrical particle as an infinitely long cylinder with radius $b$, delimited by a free interface (\textit{i.e.} absorbing boundary condition) located at $r=b$, where $r$ is the radial coordinate of $\bm{r}$, the origin of $\bm{r}$ being chosen on the axis of the cylinder. We introduce the rescaled quantities $R=r/\xi$ and $B=b/\xi$, and the dimensionless free surface to volume ratio is $2/B$. For such a geometry, the separation of variables in the diffusion problem yields a unidimensional description along $R$, with:
\begin{equation}
u_n(R)=\frac{\sqrt{2}}{BJ_1(\omega_n B)} J_0\left(\omega_n R\right) \  ,
\end{equation}
where $J_0$ and $J_1$ are Bessel functions of the first kind, and:
\begin{equation}
\lambda_n=\frac{\omega_n^{\,2}}{2}=\frac{x_n^{\,2}}{2B^2}\ , \ \forall n\geq1\ ,
\end{equation}
with the $x_n$ for $n\geq1$ being all the positive zeros of $J_0$. Note that the integrals $\int_D\text{d}\bm{R}$ are replaced by $\int_0^B\text{d}R\,R$. With those specific spectral quantities, the cooperative reduction factor of Eq.~(\ref{coopred}) becomes:
\begin{equation}
\label{solcyl}
f(R,B)=\frac{4}{B}\sum_{n=1}^{+\infty}\frac{J_0\left(\omega_n R\right)}{J_1(\omega_n B)}\,\frac{1- e^{-\frac{\omega_n^{\,2}}{2}}}{\omega_n^{\,3}}\ ,
\end{equation}
where we have now indicated explicitly the previously hidden dependency on the geometry, through the unique confinement parameter $B$. 

Using Eq.~(\ref{solcyl}) and the procedure detailed in Sec.~\ref{princ}, we can now compute $T_{\textrm{g}}(b)$ and $h_{\textrm{m}}(T,b)$ for (low molecular weight) polystyrene, which are plotted on Figs.~\ref{Fig2} and~\ref{Fig5}, respectively. The trends are qualitatively similar to the ones for other geometries. Moreover, the $T_{\textrm{g}}(b)$ curve for cylindrical particles is quantitatively close to the $T_{\textrm{g}}(h)$ one for freestanding films, which is expected since the free surface to volume ratio is the same in both geometries, for identical $b$ and $h$ values.

\section{Effect of a purely attractive boundary}
\begin{figure}[!t]
\centering
\includegraphics[scale=0.4]{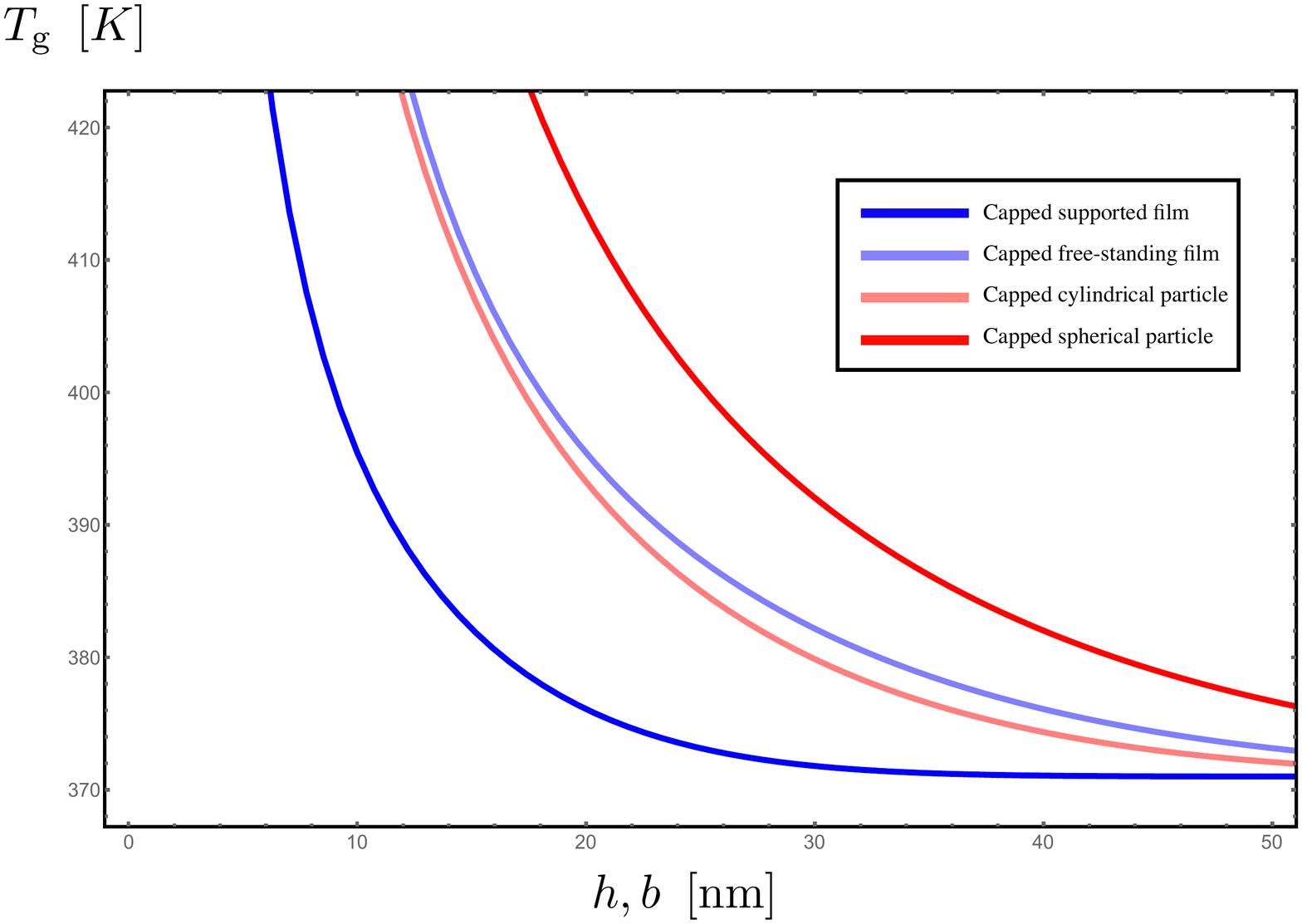}
\caption{Effective glass transition temperature of (low molecular weight) polystyrene as a function of system size (\textit{i.e.} film thickness $h$ or particle radius $b$), for various confined geometries involving free interfaces capped by purely attractive substrates, and in one case (supported film) also a passive substrate, as computed from Eqs.~(\ref{survival}),~(\ref{fpd})~and~(\ref{last}), as well as the procedure detailed in Sec.~\ref{princ}.}
\label{Fig6}
\end{figure}

Up to now we have addressed free interfaces (\textit{i.e.} absorbing boundary condition), and passive substrates (\textit{i.e.} reflecting boundary condition), or mixtures of the two, within the context of the cooperative string model. Through this last section, we would like to extend the range of applicability of the model to more realistic substrates, that can attract the sample molecules and inhibit their cooperative motion. To do so, we invoke a purely attractive boundary condition. A cooperative string that reaches this boundary with a number of steps $n=lN^*$ smaller than the bulk cooperativity $N^*$ does not contribute to relaxation. Nevertheless, the test particle (see Fig.~\ref{Fig1}) may still relax in the following ways. First, relaxation can occur through the bulk mechanism with the (survival) probability $S(\mathbf{R},1)$ (see Eq.~(\ref{survival})) and a cooperativity $N^*$. Secondly, if the cooperative string touches the attractive boundary with $l<1$, which happens with probability $p(\mathbf{R})=1-S(\mathbf{R},1)$, and an average number of steps $\bar{l}N^*$, where we introduced the conditional average $\bar{l}(\mathbf{R})=\int_0^{1}\textrm{d}l_0\,g(\bm{R},l_0)$ (see Eq.~(\ref{fpd})), the realization does not succeed, as explained above, but a second attempt can be performed. Then, the probability of success becomes $p(1-p)$, and in case of success the average cooperativity is $\left(1+\bar{l}\right)N^*$ by addition of the respective values for the two attempts. By iterating the process, the probability of relaxation at the $k$-th attempt is $p^{k-1}(1-p)$, and in case of success the average cooperativity is $[1+(k-1)\bar{l}]N^*$. Finally, from a partition on all possible cases, the average local cooperativity reads $N_{\textrm{s}}^*=N^*f$, with the cooperative reduction factor:
\begin{equation}
f(\mathbf{R})=\sum_{k=1}^{+\infty}p(\mathbf{R})^{k-1}\left[1-p(\mathbf{R})\right]\left[1+(k-1)\bar{l}(\mathbf{R})\,\right]\ .
\end{equation}
After calculation of the sum, this leads to:
\begin{equation}
\label{last}
f(\mathbf{R})=1+\frac{1-S(\mathbf{R},1)}{S(\mathbf{R},1)}\,\bar{l}(\mathbf{R})\ .
\end{equation}
Using Eqs.~(\ref{survival}),~(\ref{fpd})~and~(\ref{last}), as well as the procedure detailed in Sec.~\ref{princ}, one can compute the effective glass transition $T_{\textrm{g}}$ as function of the system size $h$ or $b$, in various situations where the free interfaces (\textit{i.e.} absorbing boundary condition) are capped by purely attractive substrates (\textit{i.e.} purely attractive boundary condition). The results for (low molecular weight) polystyrene capped freestanding films, capped supported films, capped spherical particles, and capped cylindrical particles, are shown in Fig.~\ref{Fig6}. Interestingly, we observe an increase of $T_{\textrm{g}}$ with decreasing sample size -- an effect which is more marked for larger capped surface to volume ratios. The results of Figs.~\ref{Fig2} and~\ref{Fig6} taken together are thus reminiscent of the unified zoology of apparently contradictory observations made on different substrates/interfaces~\cite{Torres2000}.

\section{Conclusion}
In this article, we have explored some key novel aspects of the previously introduced cooperative string model for confined glasses. In particular, we have studied the effective glass transition and surface mobility of various experimentally-relevant nanoconfined geometries, involving free interfaces and passive substrates: freestanding films, supported films, spherical particles, and cylindrical particles. Finally, we have extended the range of applicability of the model to purely attractive substrates, and have explored the impact of the latter in the previous geometries. As such, the cooperative string model allows to recover the various observations made for glassy polystyrene in the literature. These important and diverse validated features of the model, combined to the analytical simplicity, confirm its practical interest for the study of the glass transition in confinement and at interfaces. Other situations, such as partially attractive walls, liquid and rubbery substrates, glassy binary mixtures, inclusions, macromolecular polymeric effects, fragility, or ageing, to name a few, should thus be studied with the cooperative string model in future.

\begin{acknowledgments}
The authors thank Justin Salez, Kari Dalnoki-Veress and Ulysse Mizrahi for interesting discussions. They acknowledge financial support from the Paris-Sciences and Joliot Chairs of ESPCI Paris, and from the Perimeter Institute for Theoretical Physics. Research at Perimeter Institute is supported by the Government of Canada through Industry Canada and by the Province of Ontario through the Ministry of Economic Development $\&$ Innovation.
\end{acknowledgments}
\bibliographystyle{unsrt}
\bibliography{Arutkin2019}
\end{document}